\documentclass{chi-ext}

\title{Learning styles: Literature versus machine learning}

\numberofauthors{3}
\author{
  \alignauthor{
  	\textbf{Farah Bouassida}\\
  	\affaddr{
\'Ecole Polytechnique F\'ed\'erale
de Lausanne
}\\
  	\affaddr{RLC D1 740, CH-1015, Lausanne, Switzerland}\\
  	\email{farah.bouassida@epfl.ch }
  }
  \vfil
  \alignauthor{
  	\textbf{{\L}ukasz Kidzi\'nski}\\
  	\affaddr{
\'Ecole Polytechnique F\'ed\'erale
de Lausanne}\\
  	\affaddr{RLC D1 740, CH-1015, Lausanne, Switzerland}\\
  	\email{lukasz.kidzinski@epfl.ch }
  }
  \vfil
    \alignauthor{
  	\textbf{Pierre Dillenbourg}\\
  	\affaddr{
\'Ecole Polytechnique F\'ed\'erale
de Lausanne}\\
  	\affaddr{RLC D1 740, CH-1015, Lausanne, Switzerland}\\
  	\email{pierre.dillenbourg@epfl.ch }
  }
}

\def\plaintitle{CHI LaTeX Extended Abstracts Template}
\def\plainauthor{Farah Bouassida}
\def\plainkeywords{learning styles, adaptive learning, intelligent tutoring systems}
\def\plaingeneralterms{educational data mining, learning analytics}

\hypersetup{
  pdftitle={\plaintitle},
  pdfauthor={\plainauthor},  
  pdfkeywords={\plainkeywords},
  pdfsubject={\plaingeneralterms},
}

\usepackage{graphicx}   
\usepackage{balance}    
\usepackage{bibspacing} 
\usepackage{tabularx} 

\begin{document}

\maketitle

\begin{abstract}
Every teacher understands that different students benefit from different activities. Recent advances in data processing allow us to detect and use behavioral variability for adapting to a student. This approach allows us to optimize learning process but does not focus on understanding it. Conversely, classical findings in educational sciences allow us to understand the learner but are hard to embed in a large scale adaptive system. In this study we design and build a framework to investigate when the two approaches coincide.

\end{abstract}

\keywords{\plainkeywords, \plaingeneralterms}

\category{K.3.1}{Computer Uses in Education}{Computer-managed instruction (CMI)}.  

\section{Introduction}

The spread of Massive Online Open Courses (MOOCs) and global learning environments like Khan academy changed the educational research irreversibly. More data allows researchers and practitioners to design and build tailored learning experience. However, it has also broaden the gap between classical educational science and data-driven educational research. We aim to construct a unifying framework. In this study, we approach two research questions in the context of adaptive learning:
\begin{enumerate}
\item[Q1] Do student features derived through data mining coincide with findings from educational studies?
\item[Q2] Can we suggest learning activities based solely on features derived through machine learning? 
\end{enumerate}

\section{Learning adapted to styles}
For today, the approaches to adaptive learning can be summarized by two main branches, hereafter referred to as the Cronbach line and the Markov line. The first one is supported by classical educational research and statistical methods, whereas the latter corresponds to the black-box data mining approach. We use these two names as flags that represent a certain approach to adaptation, even if we do not focus for instance on Markov chains.

\subsection{Cronbach line}
Research in education gave rise to so-called 'aptitude-treatment interaction' laws \cite{cronbach1977aptitudes}, which specify that different learners benefit from different pedagogical strategies. We distinguish, among others, high-aptitude versus low-aptitude learners, highly motivated versus poorly motivated, anxious versus self-confident, field-dependent versus field independent learners, or surface versus deep learners. All these categorisations are based on experimental results, but the variety of factors hinders the development of unifying applications.

\subsection{Markov line}
Conversely, Markov line takes into account the whole variety of features. Researchers skip the interpretation step and build models which optimize the learning gain, given the features of a student. Since it is impossible to adapt independently to every single student, we define so-called 'learning profiles' - groups of students with similar characteristics. Now, when we know how a given profile reacts to certain activities, for a new student of that profile we can suggest the best set of activities.

\begin{figure}
   \includegraphics[width=0.48\textwidth]{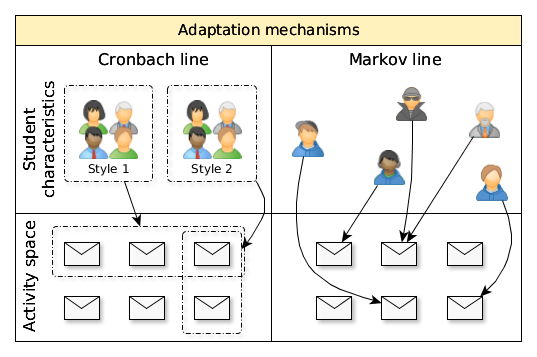}
   \caption{\label{img:markov} Two main approaches to adaptation}
\end{figure}

\section{Related work}

The idea of computer-supported adaptive education dates back to early research in artificial intelligence (AI). Several AI approaches were successfully applied, among others: 1) \textbf{Recommender systems}, systems which for given set of features and actions suggest learning materials, 2) \textbf{Adaptive systems}, a general branch where activities are adapted to a given student, 3) \textbf{Evolutive systems}, learning environments which change in time during the learning, depending on actions of a student \cite{o1979self}. For an extensive survey of data mining techniques used in educational applications we refer to \cite{romero2010educational}.

\section{Experiment}
\begin{table*}[!htbp]
\centering
    \begin{tabular}{ | p{2.5cm} | p{12cm} |}
    \hline
    Feature& Description \\ \hline
    IQ & Two questions derived from IQ tests. Each scored $-1$ (correct) and $-1$ (incorrect). The sum is used as the IQ feature.\\ \hline
    Logic & Five questions derived taken from the paper folding test\cite{PaperFolding}. Each scored $-1$ (correct) and $-1$ (incorrect). The sum is used as the logic feature.\\ \hline
    Conscientiousness \& Openness & For each of two concepts we ask five questions from International Personality Item Pool \cite{BigFive}. The answer vary from "I strongly disagree" to "I strongly agree" and the points range from 1 to 5 or from -1 to -5 in the case we give negative marks.\\
    \hline
    \end{tabular}
    \caption{\label{tbl:test} Psychological features used to derive learning styles.}
\end{table*}

\begin{figure}
   \includegraphics[width=0.47\textwidth]{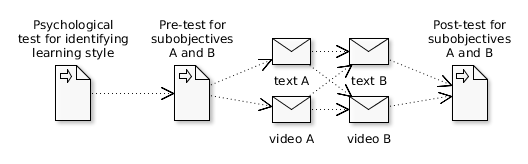}
   \caption{\label{img:experiment} The experiment design.}
\end{figure}

We design an experiment with four different learning scenarios as illustrated in Figure \ref{img:experiment}. Learning profile is determined by the psychological test described below. We design a small space of loosely correlated behavioral features, so that we can test different scenarios for similar users, with a relatively small sample of participants.

\subsection{Psychological test}

We ask questions related to different aspects of a person: psychological, demographic or IQ estimation. We list them together with the corresponding rating system in Table \ref{tbl:test}. Openness and conscientiousness are proven to be correlated to a person's motivation to learn \cite{major2006linking}, thus we hypothesise that different activities depending on values of these features can be beneficial.




For the collection of demographic information, we ask participants about their age, country, gender and the level of education ("Haven't graduated high school", "High school graduate", "Apprenticeship", "College student", "Bachelors", "Masters", "Doctorate").




\subsection{Learning activities}
We teach the concept of "epidemics" broken down to two concepts taught in the following order by two consecutive activities: A) Epidemics definition and its causes, B) epidemics reproductive number. 
An activity is either reading an encyclopedia-style text (\emph{text}) or watching a video from a MOOC\cite{EpidemicsMOOC} (\emph{video}). 

\subsection{Implementation}
We developed a web platform\cite{gitLearningPlatform} 
which executes the work-flow presented in Figure \ref{img:experiment}. Pre-test is the same as the post-test. For each student we measure the learning gain as a difference between the results from the tests. A user cannot take navigate back, but he is free to replay or navigate through a video. 

\section{Results}
We gathered a total of 167 participants but only 77 of them accomplished both pre-test and post-test. 
As our goal is to analyse the learning gain, we work with these 77 users. 
Among them, we have 25 female user and 52 male user. 
We have the following distribution for the possible learning paths: 1) \emph{text-text} (27 users), 2) \emph{text-video} (20 users), 3) \emph{video-text} (12 users) and 4) \emph{video-video} (18 users). In this study we focus on extreme unimodal scenarios: \emph{video-video} and \emph{text-text}.

\begin{figure}
\centering
\includegraphics[width=\columnwidth]{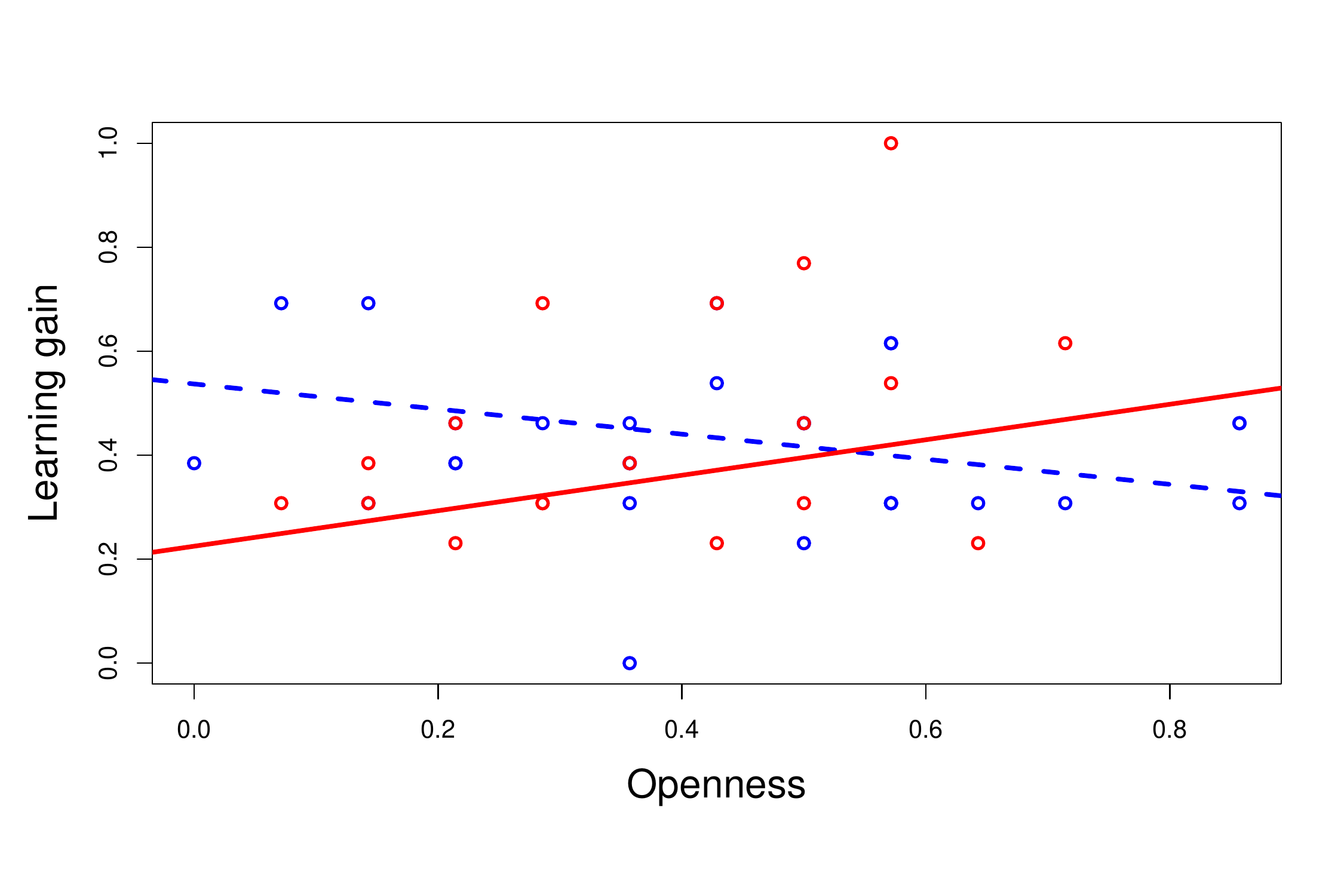}%
\caption{Linear regression for the learning gain depending on openness. The dashed line shows a decreasing tendency for text-text path while the solid line shows an increasing tendency for the video-video path.}
\label{opennessplot}
\end{figure}

In our preliminary study, we plot each collected variable against the learning gain achieved by the learner and compare it for the obtained paths. 
Figure \ref{opennessplot} shows that, in the \emph{video-video} path, the more open the learner is, the higher learning gain the learner gets. On the other hand, the \emph{text-text} path seems to have the opposite trend.
These results confirm a correlation between openness and a person's learning experience. The difference for the two learning paths shows that the openness feature could contribute to choosing the best activity for a learner. 

\section{Conclusion}
Our results are in line with studies on the openness, which indicates a partially positive answer to Q1. Such or even more implicit features can be extracted automatically with large sample at hand, what leads to a confirmatory answer to Q2. We consider these results as encouraging to continue research in this direction.

\balance
\bibliographystyle{acm-sigchi}

\end{document}